\documentclass[runningheads]{llncs}
\usepackage{graphicx}
\graphicspath{ {./images/} }

\begin{document}
%
\title{Transfer Learning and Augmentation for Word Sense Disambiguation}
%
%
\author{Harsh Kohli\inst{}\orcidID{0000-0003-1431-6025}}
\authorrunning{H. Kohli}
%
\institute{SalesKen\\
\email{harshkohli@salesken.ai}}
\maketitle              
\begin{abstract}
Many downstream NLP tasks have shown significant improvement through continual pre-training, transfer learning and multi-task learning. State-of-the-art approaches in Word Sense Disambiguation today benefit from some of these approaches in conjunction with information sources such as semantic relationships and gloss definitions contained within WordNet. Our work builds upon these systems and uses data augmentation along with extensive pre-training on various different NLP tasks and datasets. Our transfer learning and augmentation pipeline achieves state-of-the-art single model performance in WSD and is at par with the best ensemble results.

\end{abstract}

\section{Introduction}

Word Sense Disambiguation or WSD is the task of gleaning the correct sense of an ambiguous word given the context in which it was used. It is a well-studied problem in NLP and has seen several diversified approaches over the years including techniques leveraging Knowledge-Based Systems, Supervised learning approaches and, more recently, end-to-end deep learnt models. WSD has found application in various kinds of NLP systems such as Question Answering, IR, and Machine Translation. 

WordNet 3.0 is the most popular and widely used sense inventory that consists of over 109k synonym sets or synsets and relationships between them such as hypernym, anotnym, hyponym, entailment etc. Most training and evaluation corpora used in supervised systems today consist of sentences where words are manually annotated and mapped to a particular synset in WordNet. We use these sources in addition to other publicly available datasets to tune our model for this task. Through transfer learning from these datasets and other augmentation and pre-processing techniques we achieve state-of-the-art results on standard benchmarks.

\section{Related Work}

Traditional approaches to WSD relied primarily on Knowledge-Based Systems. Lexical similarity over dictionary definitions or Gloss for each synset was first used in \cite{Lesk1986AutomaticSD} to estimate the correct sense. Graph based approaches such as \cite{Moro2014EntityLM} were also proposed which leverage structural properties of lexico-semantic sources treating the the knowledge graph as a semantic network. One major advantage of using such unsupervised techniques was that they eliminated the need of having large annotated training corpora. Since annotation is expensive given the large number of fine-grained word senses, such methods were the de facto choice for WSD systems. Recently, however, approaches for semi-automatic \cite{taghipour-ng-2015-one} and automatic \cite{pasini} sense annotation have been proposed to partially circumvent the problem of manually annotating a sizeable training set.

Supervised methods, on the other hand, relied on a variety of hand-crafted features such as a neighbouring window of words and their corresponding part of speech (POS) tags etc. Commonly referred to as word expert systems, they involved training a dedicated classifier for each individual lemma \cite{zhong-ng-2010-makes}. The default or first sense was usually returned when the target lemma was not seen during training. While these were less practical in real application, they often yielded better results on common evaluation sets.

\cite{kageback-salomonsson-2016-word} and \cite{raganato-etal-2017-neural} were the first neural architectures for WSD which consisted of Bidirectional LSTM models and Seq2Seq Encoder-Decoder architectures with attention. These architectures optionally included lexical and POS features which yielded better results. Due to strong performance of contextual embeddings such as BERT \cite{Devlin2019BERTPO} on various NLP tasks, recent approaches such as \cite{DBLP:journals/corr/abs-1905-05677} and \cite{Huang2019GlossBERTBF} have used these to achieve significant gains in WSD benchmarks. We leverage the ideas presented in GlossBERT \cite{Huang2019GlossBERTBF} and improve upon the results with a multi-task pre-training procedure and greater semantic variations in the train dataset through augmentation techniques.

\section{Data Preparation Pipeline}

\subsection{Source Datasets}

We use the largest manually annotated WSD corpus SemCor 3.0 \cite{miller-etal-1993-semantic} consisting of over 226k sense tags for training our models. In keeping with most neural architectures today such as \cite{luo-etal-2018-leveraging}, we use the SemEval-2007 corpus \cite{pradhan-etal-2007-semeval} as our dev set and SemEval-2013 \cite{navigli-etal-2013-semeval}, SemEval-2015\cite{moro-navigli-2015-semeval}, Senseval-2 \cite{edmonds-cotton-2001-senseval}, and Senseval-3 \cite{snyder-palmer-2004-english} as our test sets.

\subsection{Data Preprocessing}

GlossBERT \cite{Huang2019GlossBERTBF} utilizes context gloss pairs with weak supervision to achieve state-of-the-art single model performance on the evaluation sets. We follow the same pre-processing procedure as GlossBERT. The context sentence along with each of the gloss definitions of senses of the target word are considered as a pair. Thus, for a sentence containing an ambiguous word with N senses, we consider all N senses with as many sentence pairs. Only the correct sense is marked as a positive sample while all others are considered negative inputs to our pairwise sentence classifier. As this formulation relies on the gloss definition of a synset and not just the synset tag or key, it is more robust to keys that do not occur or are under-represented in training.

\begin{figure*}
 \center
  \includegraphics[width=\textwidth, height=4cm]{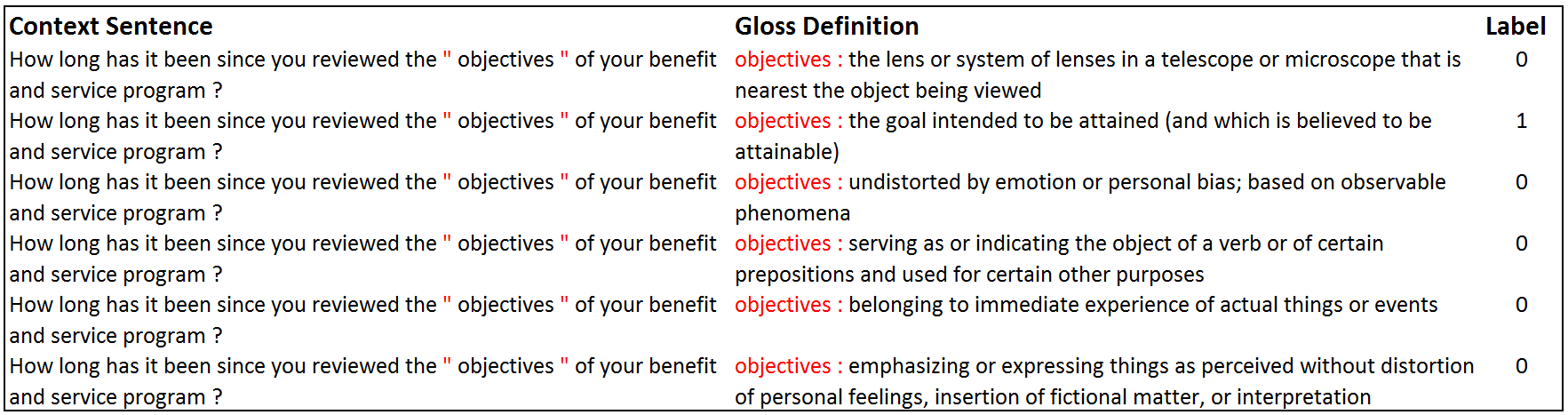}
  \caption{Context-Gloss Pairs with Weak Supervision}
  \label{fig:sample}
\end{figure*}

Figure ~\ref{fig:sample} above shows an example of context-gloss pairs for a single context sentence with the target word - objectives. The highlighted text represent the weak supervised signals which help identify the target word both in the gloss definition, as well as in the context sentence. In the context sentence, the target word may appear more than once, and the signal helps associate each occurrence with the definition independently. 

\subsection{Data Augmentation}

Given the large number of candidate synsets for each target lemma, the train dataset has a large class imbalance. The ratio of negative samples to positives is nearly 8:1. Rather than adopting a simple oversampling strategy, we use data augmentation through back translation. Back translation is a popular method for generating paraphrases involving translating a source sentence to one of several target languages and then translating the sentence back into the source language. Approaches described in \cite{prakash-etal-2016-neural}, \cite{mallinson-etal-2017-paraphrasing}, \cite{wieting-gimpel-2018-paranmt} have successfully leveraged modern Neural Machine Translation systems to generate paraphrases for a variety of tasks. We use this technique to introduce greater diversity and semantic variation in our training set and augment examples in our minority class. 

The Transformers library \cite{Wolf2019HuggingFacesTS} provides MarianMT models \cite{mariannmt} for translation to and from several different languages. Each model is a 6-layer transformer \cite{NIPS2017_7181} encoder-decoder architecture. For best results, we select from a number of high-resource languages such as French, German etc. and apply simple as well as chained back-translation (e.g. English - Spanish - English - French - English). From our pool of back-translated sentences, we retain sentences where the target word occurs exactly once in the original as well as back-translated sentence. This way, we generate several paraphrased examples for each positive example in our train set. We randomly select $n$ augmented samples for each original sample at train time, where $n$ was treated as a hyper-parameter during our training experiments. We achieve best results when $n=3$.

\section{Model}

We use the MT-DNN \cite{liu2019mt-dnn} architecture for training our model. The network consists of shared layers and task-specific layers. Through cross-task training, the authors demonstrate how the shared layers of the network learn more generalized representations and are better suited to adapt to new tasks and domains. Multi-task learning using large amount of labelled data across tasks has a regularization effect on the network and the model is able to better generalize to new domains with relatively fewer labelled training examples than simple pre-trained BERT. It is this property of MT-DNN that we leverage to improve performance on WSD.

\begin{figure*}
 \center
  \includegraphics[width=\textwidth, height=5cm]{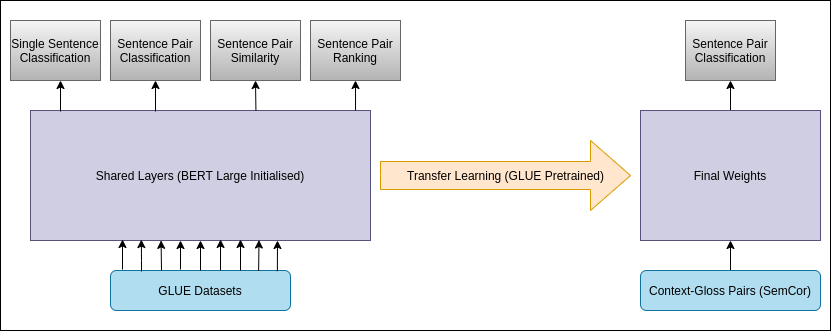}
  \caption{Pre-training and Tuning methodology}
  \label{fig:modelDiagram}
\end{figure*}

The pre-training procedure for MT-DNN is similar to that of BERT which used two supervised tasks - masked LM and next sentence prediction. Using BERT Large model (24 layers, 1024 dim, 335m trainable parameters) as our base model, we then tune on all tasks in the GLUE benchmark \cite{wang-etal-2018-glue}. While \cite{Huang2019GlossBERTBF} reported better performance using BERT base (12 layers, 768 dim, 110m trainable parameters), we found that the larger BERT model performed significantly better in our experiments. We attribute this behaviour to our pre-training procedure which learns better, more generalized representations thus preventing a larger, more expressive model from overfitting on the train dataset. 

Four different task-specific output layers are constructed corresponding to single sentence classification, pairwise text similarity, pairwise text classification, and pairwise text ranking. These are illustrated in Figure ~\ref{fig:modelDiagram}. Learning objectives differ for each task - single-sentence and pairwise classification tasks are optimized using cross-entropy loss, pairwise text similarity is optimized on the mean squared error between the target similarity value and semantic representations of each of the sentences in the input pair, and pairwise text ranking follows the pairwise learning-to-rank paradigm in minimizing the negative log likelihood of a positive example given a list of candidates \cite{10.1145/1102351.1102363}. The pairwise text classification output layer uses a stochastic answer network (SAN) \cite{sannetwork} which maintains a memory state and employs K-step reasoning to iteratively improve upon predictions. We use the same pairwise classification head when tuning the network for our WSD task. At inference time, we run context-gloss pairs for each sense of the target lemma and the candidate synset with the highest score is considered the predicted sense.

\section{Implementation Details}

Examples from each of the 9 datasets in GLUE are input to the network and passed to the correct output layer given the task-type. 5 epochs of pre-training are thus carried out using GLUE data. The best saved checkpoint is then selected and, thereafter, context-gloss pairs as described above are input to the model for tuning on WSD. Model weights of shared layers are carried over from multi-task training on GLUE. Adamax \cite{adamax} optimizer is used to tune the weights and a low learning rate of 2e-5 is used to facilitate a slower, but smoother convergence. A batch size of 256 is maintained and the architecture is tuned on 8x Tesla V100 GPU's with 16GB of VRAM each for a total of 128GB GPU memory.

\section{Results}

\begin{table*}[ht]
\begin{center}
\begin{tabular}{|c|c|c|c|c|c|c|c|c|c|c|} 
\hline \bf System & \bf SE07 & \bf SE2 & \bf SE3 & \bf SE13 & \bf SE15 & \bf Noun & \bf Verb & \bf Adj & \bf Adv & \bf All\\ \hline
MFS Baseline & 54.5 & 65.6 & 66.0 & 63.8 & 67.1 & 67.7 & 49.8 & 73.1 & 80.5 & 65.5\\
Lesk\textsubscript{ext+emb} & 56.7 & 63.0 & 63.7 & 66.2 & 64.6 & 70.0 & 51.1 & 51.7 & 80.6 & 64.2\\
Babelfly & 51.6 & 67.0 & 63.5 & 66.4 & 70.3 & 68.9 & 50.7 & 73.2 & 79.8 & 66.4\\
IMS & 61.3 & 70.9 & 69.3 & 65.3 & 69.5 & 70.5 & 55.8 & 75.6 & 82.9 & 68.9\\
IMS\textsubscript{+emb} & 62.6 & 72.2 & 70.4 & 65.9 & 71.5 & 71.9 & 56.6 & 75.9 & 84.7 & 70.1\\
Bi-LSTM & - & 71.1 & 68.4 & 64.8 & 68.3 & 69.5 & 55.9 & 76.2 & 82.4 & 68.4\\
Bi-LSTM\textsubscript{+att.+LEX+POS} & 64.8 & 72.0 & 69.1 & 66.9 & 71.5 & 71.5 & 57.5 & 75.0 & 83.8 & 69.9\\
GAS\textsubscript{ext}(Linear) & - & 72.4 & 70.1 & 67.1 & 72.1 & 71.9 & 58.1 & 76.4 & 84.7 & 70.4\\
GAS\textsubscript{ext}(Concatenation) & - & 72.2 & 70.5 & 67.2 & 72.6 & 72.2 & 57.7 & 76.6 & 85.0 & 70.6\\
CAN & - & 72.2 & 70.2 & 69.1 & 72.2 & 73.5 & 56.5 & 76.6 & 80.3 & 70.9\\
HCAN & - & 72.8 & 70.3 & 68.5 & 72.8 & 72.7 & 58.2 & 77.4 & 84.1 & 71.1\\
SemCor,hyp & - & - & - & - & - & - & - & - & - & 75.6\\
SemCor,hyp(ens)* & 69.5 & 77.5 & 77.4 & 76.0 & 78.3 & 79.6 & 65.9 & 79.5 & 85.5 & 76.7\\
SemCor+WNGC,hyp & - & - & - & - & - & - & - & - & - & 77.1\\
SemCor+WNGC,hyp(ens)* & 73.4 & 79.7 & 77.8 & 78.7 & 82.6 & 81.4 & 68.7 & 83.7 & 85.5 & 79.0\\
BERT(Token-CLS) & 61.1 & 69.7 & 69.4 & 65.8 & 69.5 & 70.5 & 57.1 & 71.6 & 83.5 & 68.6\\
GlossBERT(Sent-CLS) & 69.2 & 76.5 & 73.4 & 75.1 & 79.5 & 78.3 & 64.8 & 77.6 & 83.8 & 75.8\\
GlossBERT(Token-CLS) & 71.9 & 77.0 & 75.4 & 74.6 & 79.3 & 78.3 & 66.5 & 78.6 & 84.4 & 76.3\\
GlossBERT(Sent-CLS-WS) & 72.5 & 77.7 & 75.2 & 76.1 & 80.4 & 79.3 & 66.9 & 78.2 & 86.4 & 77.0\\
MTDNN+Gloss & \bf 73.9 & \bf 79.5 & \bf 76.6 & \bf 79.7 & \bf 80.9 & \bf 81.8 & \bf 67.7 & \bf 79.8 & \bf 86.5 & \bf 79.0\\
\hline
\end{tabular}
\end{center}
\caption{Final Results. * Result excluded from consideration as it uses an ensemble}
\label{table:results} 
\end{table*}

We summarize the results of our experiments in Table~\ref{table:results}. We compare our results against the Most Frequent Sense Baseline as well as different approaches, Knowledge Based - Lesk (ext+emb) \cite{basile-etal-2014-enhanced} and Babelfly \cite{Moro2014EntityLM}, Word-Expert Supervised Systems - IMS \cite{zhong-ng-2010-makes} and IMS+emb\cite{iacobacci-etal-2016-embeddings}, Neural Models - Bi-LSTM \cite{kageback-salomonsson-2016-word}, Bi-LSTM + att + lex +pos \cite{raganato-etal-2017-neural}, CAN/HCAN \cite{luo-etal-2018-leveraging}, GAS \cite{luo-etal-2018-incorporating}, SemCar/SemCor+WNGC, hypernyms \cite{DBLP:journals/corr/abs-1905-05677} and GlossBERT \cite{Huang2019GlossBERTBF}. We exclude results from ensemble systems marked in Table ~\ref{table:results} as these results were obtained using a geometric mean of predictions across 8 independent models. We achieve the best results for any single model across all evaluation sets and POS types. 

While \cite{DBLP:journals/corr/abs-1905-05677} supplement their train corpus with the Wordnet Gloss Corpus (WNGC) and also use 8 different models for their ensemble, our overall results are at par with theirs on test datasets and slightly better on the dev set. The fact that such results were achieved with fewer training examples (without the use of WNGC) further enforces the generalization and domain adaptation capabilities of our pre-training methodology.

\section{Conclusion \& Future Work}

We use the pre-processing steps and weak-supervision over context-gloss pairs as described in \cite{Huang2019GlossBERTBF} and improve upon the results through simple and chained back-translation as a means of data augmentation and multi-task training and transfer learning from different data sources. Better and more generalized representations achieved by leveraging the GLUE datasets allows us to train a larger model with nearly thrice as many trainable parameters. Through these techniques we are able improve upon existing SOTA on standard benchmark.

Additional data from WNGC or OMSTI \cite{taghipour-ng-2015-one} has shown to aid model performance in various systems and could be incorporated in training. Recent work such as \cite{tayyar-madabushi-etal-2019-cost} indicates that cost-sensitive training is often effective when training BERT when there is a class imbalance. Given the nature of the problem, a triplet loss function similar to \cite{7298682} could be used to further improve performance. Online hard or semi-hard sampling strategies could be experimented with to sample the negative sysnets. Finally, RoBERTa \cite{DBLP:journals/corr/abs-1907-11692} has shown improved performance on many NLP tasks and could be used as a base model that is input to our multi-task pre-training pipeline. All of these techniques could be used in conjunction with our context-gloss pairwise formulation to improve performance further.

%
%
%
\bibliographystyle{splncs04}
\bibliography{mybibliography}
%

\end{document}